\documentclass[twocolumn,10pt,a4paper]{article}

\usepackage[utf8]{inputenc}
\usepackage[T1]{fontenc}
\usepackage[english]{babel}
\usepackage{geometry}
\usepackage{amsmath,amssymb}
\usepackage{booktabs}
\usepackage{tabularx}
\usepackage{multirow}
\usepackage{graphicx}
\usepackage{xcolor}
\usepackage{hyperref}
\usepackage{caption}
\usepackage{pgfplots}
\pgfplotsset{compat=1.18}
\usepgfplotslibrary{groupplots}

\hypersetup{
	colorlinks=true,
	linkcolor=blue!60!black,
	citecolor=blue!60!black,
	urlcolor=blue!60!black
}

\title{\textbf{Multi-Site Astronomical, Meteorological, and Geological Comparative Study across Italian Localities (1990--2014): The NGICS Synthesis}}

\author{
	\begin{tabular}{ccc}
		Diego Tasselli$^{1,4}$ & Stefano Ricci$^{2,4}$ & Pamela Bianchi$^{3,4}$ \\
		\footnotesize \texttt{diego.tasselli@tscorporation.it} & \footnotesize \texttt{stefano.ricci@tscorporation.it} & \footnotesize \texttt{pamela.bianchi@tscorporation.it}
	\end{tabular} \\[0.5cm]
	\small $^1$TS Corporation -- Astronomy and Astrophysics Department, Via Rugantino 71, 00169 Roma RM, Italy \\
	\small $^2$TS Corporation -- Meteorological and Climatic Change Department, Via Giordano Bruno 3, 73057 Taviano LE, Italy \\
	\small $^3$TS Corporation -- Geology and Geophysics Department, Via Reclusorio 38, 06084 Foligno PG, Italy \\
	\small $^4$Halminak S.r.l., Parent Company and Strategic Coordination Node, Via Rugantino 71, 00169 Roma RM, Italy
}

\date{}
\begin{document}
	\twocolumn[
	\begin{@twocolumnfalse}
		\maketitle
		\begin{abstract}
			This comparative synthesis paper, developed within the framework of the \textbf{NGICS (New Italian City Geological Study)} and \textbf{Project Sismes} programs, aggregates and cross-examines 25 years (1990--2014) of multidisciplinary environmental, climatic, hydrological, and geological datasets across selected Italian geographic sectors: \textbf{Acquapendente (VT)}, \textbf{Montefiascone (VT)}, \textbf{Pieve a Nievole (PT)}, \textbf{Taviano (LE)}, and the \textbf{Torrespaccata District of Rome (RM)}. Following the analytical protocol established for the Torrespaccata District, every raw micrometeorological parameter, solar irradiation index, hygrometric/barometric series, anemometric profile, and stratigraphic-seismological record was extracted and homogenized across all five municipalities. Statistical comparison reveals a coherent macroclimatic transition characterized by rising thermal extremes, localized shifts toward high-intensity convective rainfall regimes, declining radiation fog frequency due to nocturnal warming, and distinct seismotectonic and hydrogeological vulnerabilities governed by regional volcanic, graben, and carbonate substrates.
			\vspace{0.25cm}
			
			\vspace{0.5cm}
		\end{abstract}
	\end{@twocolumnfalse}
	]

	\section{Introduction}
	Long-term environmental monitoring across heterogeneous geographic contexts is essential to decouple global macroclimatic trends from localized microclimatic, topographic, and geological forcing factors. This synthesis consolidates the empirical findings of five individual site-testing investigations conducted by TS Corporation over the 1990--2014 baseline \cite{tasselli2015acq,tasselli2015tav,tasselli2019pie,tasselli2021mon,tasselli2026tor}. Modeling its structure on the \textbf{Torrespaccata District (Rome)} protocol \cite{tasselli2026tor}, this paper provides a complete parameter-by-parameter comparison across all five localities.
	
	\subsection{Geographic, Topographic and Regulatory Framework}
	The five investigated sites span central-northern, central-Tyrrhenian, and southern-Ionian Italy, encompassing coastal plains, alluvial basins, volcanic calderas, and dense metropolitan suburbs (Table~\ref{tab:geo_overview}).
	
	\begin{table*}[t]
		\centering
		\caption{Geographic, topographic, climatic (D.P.R. 412/1993), and seismic classification of the five investigated localities.}
		\label{tab:geo_overview}
		\footnotesize
		\setlength{\tabcolsep}{4pt}
		\resizebox{\textwidth}{!}{%
			\begin{tabular}{lccccc}
				\toprule
				\textbf{Locality (Province)} & \textbf{Coordinates} & \textbf{Elevation (m a.s.l.)} & \textbf{Geomorphological Setting} & \textbf{Climatic Zone (HDD)} & \textbf{Seismic Zone ($a_g$)} \\
				\midrule
				\textbf{Acquapendente (VT)} & $42^\circ44'41.28''\text{N}$, $11^\circ51'54.36''\text{E}$ & $\sim 420$ (Sub-valley) & Paglia basin / Volcanic plateau & Zone E (2,299) & 2B ($< 0.20\text{g}$) \\
				\textbf{Montefiascone (VT)} & $42^\circ32'25.08''\text{N}$, $12^\circ02'13.56''\text{E}$ & $590$ & Southern rim Lake Bolsena caldera & Zone E (2,467) & 2B ($< 0.20\text{g}$) \\
				\textbf{Pieve a Nievole (PT)} & $43^\circ52'21.72''\text{N}$, $10^\circ48'09.00''\text{E}$ & $14 - 254$ & Valdinievole plain / Fucecchio swamp & Zone D (1,708) & 3 ($< 0.15\text{g}$) \\
				\textbf{Taviano (LE)} & $39^\circ59'04.20''\text{N}$, $18^\circ05'15.36''\text{E}$ & $57$ & Southern Salento / Ionian coast & Zone C (1,099) & 4 ($< 0.05\text{g}$) \\
				\textbf{Torrespaccata (RM)} & $41^\circ54'29.34''\text{N}$, $12^\circ28'54.48''\text{E}$ & $50 - 70$ ($60$ avg) & Eastern Rome ignimbritic plateau & Zone D (1,415) & 2B ($< 0.20\text{g}$) \\
				\bottomrule
			\end{tabular}%
		}
	\end{table*}
	
	\subsection{Astronomical Site-Testing and Observing Quality Assessment}
	Although the earliest NGICS baselines~\cite{tasselli2015acq,tasselli2015tav} focused primarily on micrometeorological and geological characterization, integrating the full 1990--2014 atmospheric datasets enables a quantitative \textit{astronomical site-testing} evaluation across all five localities (Table~\ref{tab:astronomy}). Astronomical observing efficiency is governed by the interplay of cloud-free night fraction, atmospheric water vapor/aerosol transparency, boundary-layer thermal stability (seeing), and artificial light pollution (Bortle scale):
	
	\begin{itemize}
		\item \textbf{Taviano (LE):} Exhibits the highest annual clear-night fraction ($\sim 260\text{--}270\text{ nights/yr}$) and maximum solar transparency (mean DNI $6,040\text{ Wh/m}^2\text{/day}$), making it well suited for continuous time-series CCD photometry, exoplanet transit monitoring, and solar astrophysics, despite moderate coastal humidity ($69.38\%$).
		\item \textbf{Acquapendente (VT) \& Montefiascone (VT):} Benefit from darker rural-to-suburban skies (Bortle Class~4--5) across the Vulsini plateau. Montefiascone's $590\text{ m}$ elevation reduces low-level aerosol extinction, though winter Tramontana gales (up to $100\text{ km/h}$) and Lake Bolsena humidification fogs ($923\text{ days}$) constrain winter observing windows. Acquapendente offers dark skies suitable for faint-target CCD photometry, punctuated by nocturnal valley thermal inversions.
		\item \textbf{Pieve a Nievole (PT):} Located inside the Valdinievole semi-graben, its observing duty cycle ($\sim 210\text{--}220\text{ nights/yr}$) is limited by persistent autumn radiation fogs ($1,052\text{ days}$) and high relative humidity ($74.17\%$), favoring bright variable-star photometry and daytime solar observations.
		\item \textbf{Torrespaccata District (RM):} Despite a high number of clear nights ($\sim 240\text{--}250\text{ nights/yr}$) and declining fog frequency ($254\text{ days}$) driven by the Urban Heat Island (UHI), severe metropolitan light pollution (Bortle Class~8--9) and nocturnal convective thermal plumes restrict optical observations to bright-star photometry, narrowband emission-line imaging, and meteor/radio monitoring.
	\end{itemize}
	
\begin{table*}[htbp]
	\centering
	\caption{Comparative astronomical site-testing parameters derived from the 1990--2014 NGICS atmospheric baselines.}
	\label{tab:astronomy}
	\resizebox{\textwidth}{!}{%
		\begin{tabular}{lcccc}
			\hline
			\textbf{Locality (Province)} & \textbf{Est. Clear Nights/yr} & \textbf{Boundary-Layer Seeing Regime} & \textbf{Sky Quality (Bortle)} & \textbf{Primary Astronomical Suitability} \\
			\hline
			Acquapendente (VT) & $235\text{--}245$ & Moderate--Good (valley inversions) & Class 4 & Deep-sky CCD photometry, variable stars \\
			Montefiascone (VT) & $225\text{--}235$ & Good--Variable ($590\text{ m}$ rim; winter gales) & Class 4--5 & High-altitude photometry, planetary/solar \\
			Pieve a Nievole (PT) & $210\text{--}220$ & Moderate (basin inversions; high RH/fog) & Class 5--6 & Bright variable stars, solar monitoring \\
			Taviano (LE) & $260\text{--}270$ & Good (steady coastal flow; max DNI) & Class 4--5 & Time-series photometry, solar physics \\
			Torrespaccata (RM) & $240\text{--}250$ & Urban-perturbed (UHI convective plumes) & Class 8--9 & Narrowband imaging, bright stars, radio \\
			\hline
		\end{tabular}%
	}
\end{table*}

	\section{Meteo-Climatic Parameters (1990--2014)}
	All atmospheric variables---air temperatures ($T$), Dew Point ($T_{\text{dew}}$), Relative Humidity ($\text{RH}$), surface barometric pressure ($P_{\text{surf}}$), diurnal thermal range ($\Delta T$), precipitation regimes, anemometry, and solar irradiation---were homogenized over the 25-year operational window (1990--2014) using statistical imputation to mitigate missing-data biases. Table~\ref{tab:baseline_all} synthesizes the complete set of baseline parameters across all five sites.
	
	\begin{table*}[t]
		\centering
		\caption{Comprehensive comparative matrix of baseline meteorological and climatological parameters (1990--2014).}
		\label{tab:baseline_all}
		\footnotesize
		\setlength{\tabcolsep}{4pt}
		\resizebox{\textwidth}{!}{%
			\begin{tabular}{lcccccc}
				\toprule
				\textbf{Climatic Parameter} & \textbf{Unit} & \textbf{Acquapendente} & \textbf{Montefiascone} & \textbf{Pieve a Nievole} & \textbf{Taviano} & \textbf{Torrespaccata} \\
				\midrule
				Average annual temperature ($T_{\text{mean}}$) & $^\circ\text{C}$ & $14.33$ & $14.40$ & $14.98$ & $16.76$ & $15.81$ \\
				Warmest monthly record & $^\circ\text{C}$ & $28.27$ (Aug-03) & $39.10^*$ (Aug-03) & $27.75$ (Aug-03) & $28.16$ (Jul-12) & $27.94$ (Aug-03) \\
				Coldest monthly record & $^\circ\text{C}$ & $1.42$ (Feb-12) & $-10.00^*$ (Feb-91) & $1.55$ (Feb-12) & $4.43$ (Feb-12) & $2.98$ (Feb-12) \\
				Annual diurnal thermal range ($\Delta T$) & $^\circ\text{C}$ & $9.97$ & $9.87$ & $10.10$ & $9.90$ & $9.83$ \\
				Months with average $T > 20^\circ\text{C}$ & Count & $83$ & $199$ & $89$ & $104$ & $97$ \\
				Mean annual relative humidity ($\text{RH}$) & $\%$ & $67.56$ & $66.63$ & $74.17$ & $69.38$ & $73.94$ \\
				Mean annual dew point ($T_{\text{dew}}$) & $^\circ\text{C}$ & $8.70^\dagger$ & $8.91$ & $10.07$ & $11.80^\dagger$ & $10.73$ \\
				Mean surface atmospheric pressure ($P_{\text{surf}}$) & $\text{hPa}$ & $1013.00^\dagger$ & $1014.25$ & $1015.00$ & $1013.00^\dagger$ & $1013.00$ \\
				Total cumulative rainfall (1990--2014) & $\text{mm}$ & $21,401.93$ & $37,586.20$ & $26,252.48$ & $16,003.70$ & $32,614.60$ \\
				Mean annual precipitation ($P_{\text{yr}}$) & $\text{mm/yr}$ & $856.08$ & $1,503.45$ & $1,050.10$ & $640.15$ & $1,304.58$ \\
				Maximum monthly rainfall record & $\text{mm}$ & $285.90$ (Nov-08) & $660.00$ (Aug-95) & $637.00$ (Oct-90) & $343.00$ (Oct-04) & $970.00$ (Oct-92) \\
				Minimum monthly rainfall record & $\text{mm}$ & $1.50$ (Jun-04) & $2.50$ (May-06) & $1.00$ (Jul-09) & $0.00$ (Aug-00) & $0.00$ (Mar-94) \\
				Total rain days & Count & $1,198$ & $1,281$ & $1,868$ & $1,092$ & $1,711$ \\
				Total fog days & Count & $912$ & $923$ & $1,052$ & $770$ & $254$ \\
				Total storm / thunder days & Count & $181$ & $137$ & $98$ & $188$ & $140$ \\
				Total rain/storm days & Count & $506$ & $490$ & $899$ & $595$ & $1,069$ \\
				Total rain/snow days & Count & $21$ & $26$ & $19$ & $11$ & $10$ \\
				Total rain/fog days & Count & $159$ & $136$ & $260$ & $45$ & $38$ \\
				Total rain/thunder/fog days & Count & $68$ & $66$ & $81$ & $31$ & $23$ \\
				Total snow days & Count & $22$ & $36$ & $9$ & $11$ & $11$ \\
				Mean daily wind speed (1990--2014) & $\text{km/h}$ & $23.70$ & $18.64$ & $21.50$ & $24.52$ & $22.06$ \\
				Maximum wind speed gust peak & $\text{km/h}$ & $36.57$ (Apr-08) & $100.00$ (Mar-01)$^\S$ & $33.06$ (Mar-95) & $43.45$ (Mar-02) & $50.73$ (Mar-01)$^\ddagger$ \\
				Minimum wind speed baseline & $\text{km/h}$ & $9.00$ (Dec-11) & $1.00$ (Mar-90) & $7.00$ (Nov-07) & $6.00$ (Nov-10) & $0.00$ (Feb-07) \\
				\bottomrule
				\multicolumn{7}{p{18.5cm}}{\footnotesize $^*$Absolute intra-monthly extreme records; $^\dagger$Derived from 1990--2014 time-series plots (Figures 10--11 of source papers); $^\ddagger$Convective peak $86.00\text{ km/h}$ in Dec-01; $^\S$Raw synoptic peak in the reference Viterbo (ICAO: LIRV) archive (Mar-01 and Sep-14).}
			\end{tabular}%
		}
	\end{table*}
	\subsection{Solar Radiation of the Territory}
	Solar irradiation parameters---Direct Normal Irradiance ($\text{DNI}$) and Horizontal Plane Radiation ($H_h$)---demonstrate a clear latitudinal gradient across the Italian peninsula, modulated by Apennine cloudiness and coastal transparency (Table~\ref{tab:dni_monthly} and Figure~\ref{fig:solar_dt}, left):
	\begin{itemize}
		\item \textbf{Acquapendente (VT):} Annual mean $\text{DNI}$ is \textbf{$4,670.00\text{ Wh/m}^2/\text{day}$}, peaking in July ($7,990\text{ Wh/m}^2/\text{day}$) and reaching a minimum in December ($2,290\text{ Wh/m}^2/\text{day}$), corresponding to an EU photovoltaic potential of \textbf{$1,200\text{--}1,300\text{ kWh/1kWp}$}.
		\item \textbf{Montefiascone (VT):} Exhibits high summer insolation with annual mean $\text{DNI}$ of \textbf{$5,486.17\text{ Wh/m}^2/\text{day}$} (July peak: $8,152.80\text{ Wh/m}^2/\text{day}$) and annual mean $H_h$ of $4,135.47\text{ Wh/m}^2/\text{day}$ (June peak: $6,528.00\text{ Wh/m}^2/\text{day}$), within an EU yield range of \textbf{$1,100\text{--}1,200\text{ kWh/1kWp}$}.
		\item \textbf{Pieve a Nievole (PT):} Positioned in the northern Apennine foothills, it records lower normalized $\text{DNI}$ indices (annual sum $233.68$, peaking in July at $29.11\text{ Wh/m}^2/\text{day}$) within an EU solar potential range of \textbf{$1,100\text{--}1,200\text{ kWh/1kWp}$}.
		\item \textbf{Taviano (LE):} Displays the highest solar energy potential among all sites, with an annual mean $\text{DNI}$ of \textbf{$6,040.00\text{ Wh/m}^2/\text{day}$} (July peak: $9,410\text{ Wh/m}^2/\text{day}$; December low: $3,220\text{ Wh/m}^2/\text{day}$) and an EU yield range of \textbf{$1,350\text{--}1,400\text{ kWh/1kWp}$}.
		\item \textbf{Torrespaccata (RM):} Records an annual mean daily horizontal irradiance ($H_h$) of \textbf{$5.92\text{ kWh/m}^2/\text{day}$} (monthly normalized $\text{DNI}$ peak in July at $29.11\text{ Wh/m}^2/\text{day}$, $H_h$ peak at $23.11\text{ Wh/m}^2/\text{day}$) and an EU photovoltaic yield range of \textbf{$1,100\text{--}1,200\text{ kWh/1kWp}$}.
	\end{itemize}
	
	\begin{table*}[t]
		\centering
		\caption{Monthly Direct Normal Irradiance ($\text{DNI}$, $\text{Wh/m}^2/\text{day}$) and EU photovoltaic yield across the five sites.}
		\label{tab:dni_monthly}
		\footnotesize
		\setlength{\tabcolsep}{3.5pt}
		\resizebox{\textwidth}{!}{%
			\begin{tabular}{lccccccccccccc}
				\toprule
				\textbf{Locality} & \textbf{Jan} & \textbf{Feb} & \textbf{Mar} & \textbf{Apr} & \textbf{May} & \textbf{Jun} & \textbf{Jul} & \textbf{Aug} & \textbf{Sep} & \textbf{Oct} & \textbf{Nov} & \textbf{Dec} & \textbf{EU Yield (kWh/1kWp)} \\
				\midrule
				\textbf{Acquapendente} & $2,410$ & $3,500$ & $4,030$ & $4,800$ & $5,570$ & $6,840$ & $7,990$ & $7,080$ & $5,100$ & $3,710$ & $2,620$ & $2,290$ & $1,200 - 1,300$ \\
				\textbf{Montefiascone} & $3,698$ & $4,259$ & $5,545$ & $6,227$ & $7,634$ & $8,098$ & $8,153$ & $7,126$ & $5,827$ & $4,607$ & $3,707$ & $2,953$ & $1,100 - 1,200$ \\
				\textbf{Pieve a Nievole}$^*$ & $12.05$ & $14.32$ & $19.44$ & $21.54$ & $26.24$ & $28.69$ & $29.11$ & $24.99$ & $20.69$ & $15.27$ & $11.67$ & $9.67$ & $1,100 - 1,200$ \\
				\textbf{Taviano} & $3,310$ & $3,980$ & $5,350$ & $6,390$ & $7,570$ & $9,040$ & $9,410$ & $8,680$ & $6,470$ & $5,100$ & $3,820$ & $3,220$ & $1,350 - 1,400$ \\
				\textbf{Torrespaccata}$^*$ & $12.05$ & $14.32$ & $19.44$ & $21.54$ & $26.24$ & $28.69$ & $29.11$ & $24.99$ & $20.69$ & $15.27$ & $11.67$ & $9.67$ & $1,100 - 1,200$ \\
				\bottomrule
				\multicolumn{14}{l}{\footnotesize $^*$Reported in normalized hourly/index scale in the respective source datasets.}
			\end{tabular}%
		}
	\end{table*}
	\subsection{Air Temperature, Diurnal Range ($\Delta T$) and HDD}
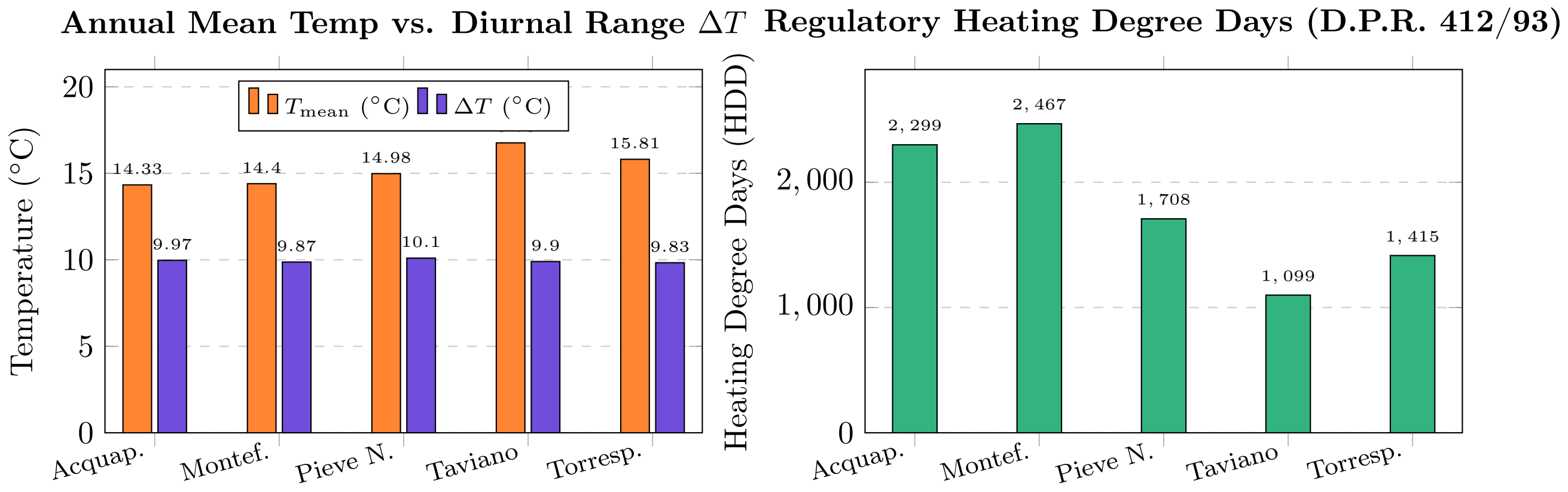
\begin{figure*}[t]
		\centering
		\begin{tikzpicture}
			\begin{groupplot}[
				group style={group size=2 by 1, horizontal sep=1.8cm},
				width=0.46\textwidth,
				height=5.6cm,
				ybar,
				symbolic x coords={Acq, Mon, Pie, Tav, Tor},
				xtick=data,
				xticklabels={Acquap., Montef., Pieve N., Taviano, Torresp.},
				xticklabel style={font=\footnotesize, rotate=15, anchor=east},
				ymajorgrids=true,
				grid style={dashed, gray!40},
				nodes near coords,
				every node near coord/.append style={font=\tiny, rotate=0, anchor=south},
				]
				\nextgroupplot[
				title={\textbf{Annual Mean Temp vs. Diurnal Range $\Delta T$}},
				ylabel={Temperature ($^\circ\text{C}$)},
				ymin=0, ymax=21,
				bar width=9pt,
				legend style={at={(0.5,0.97)}, anchor=north, legend columns=2, font=\scriptsize},
				/pgf/number format/fixed,
				/pgf/number format/precision=2
				]
				\addplot[fill=orange!80!red!80] coordinates {(Acq,14.33) (Mon,14.40) (Pie,14.98) (Tav,16.76) (Tor,15.81)};
				\addplot[fill=blue!60!violet!70] coordinates {(Acq,9.97) (Mon,9.87) (Pie,10.10) (Tav,9.90) (Tor,9.83)};
				\legend{$T_{\text{mean}}$ ($^\circ\text{C}$), $\Delta T$ ($^\circ\text{C}$)}
				
				\nextgroupplot[
				title={\textbf{Regulatory Heating Degree Days (D.P.R. 412/93)}},
				ylabel={Heating Degree Days (HDD)},
				ymin=0, ymax=2900,
				bar width=14pt,
				/pgf/number format/1000 sep={,}
				]
				\addplot[fill=teal!75!green!80] coordinates {(Acq,2299) (Mon,2467) (Pie,1708) (Tav,1099) (Tor,1415)};
			\end{groupplot}
		\end{tikzpicture}
		\caption{Comparison of annual mean temperature, diurnal thermal range $\Delta T$ (left), and regulatory Heating Degree Days (right) across the five sites.}
		\label{fig:temp_hdd}
	\end{figure*}

	Across all five localities, the 2003--2014 sub-period exhibits a marked upward thermal shift compared to 1990--2002, with \textbf{August 2003} and \textbf{July 2012} standing out as major heatwave anomalies, and \textbf{February 2012} / \textbf{February 1991} representing extreme cold spells (Figure~\ref{fig:temp_hdd}).
	
	\begin{itemize}
		\item \textbf{Thermal Inversion Dynamics:} Pronounced nocturnal radiative cooling under anticyclonic, windless conditions produces strong thermal inversions in \textbf{Acquapendente} (Paglia river valley), \textbf{Pieve a Nievole} (Valdinievole basin), and \textbf{Torrespaccata}. In Rome, winter inversions are partially counteracted by the Urban Heat Island (UHI) effect, which contracted effective Heating Degree Days from the regulatory $1,415\text{ HDD}$ down to a real 1990--2014 mean of \textbf{$1,362\text{ HDD}$} ($-3.7\%$, linear trend $-11.52\text{ HDD/year}$).
		\item \textbf{Blackbird Days Anomaly (Jan 29--31):} Evaluated in Pieve a Nievole (multi-year mean $2.22^\circ\text{C}$, showing a steady upward warming trend) and Torrespaccata (where $\Delta T_{\text{blackbird}} = \overline{T}_{\text{jan\_core}} - \mu_{\text{jan\_clim}}$ frequently exceeds the $7.56^\circ\text{C}$ January baseline, peaking at $+8.53^\circ\text{C}$ in 2002 and $+6.67^\circ\text{C}$ in 2014).
	\end{itemize}
	
	\subsection{Hygrometry, Dew Point and Atmospheric Pressure}
	\begin{itemize}
		\item \textbf{Relative Humidity ($\text{RH}$):} Evaluated via the normalized ratio $RH_{\text{ratio}} = AV/AM$, annual mean $\text{RH}$ is highest in \textbf{Pieve a Nievole} ($74.17\%$) and \textbf{Torrespaccata} ($73.94\%$, ranging from $63.59\%$ in July to $83.06\%$ in November), followed by \textbf{Taviano} ($69.38\%$, peaking in 1994 and dipping in 2008), \textbf{Acquapendente} ($67.56\%$), and \textbf{Montefiascone} ($66.63\%$, peaking in 2002 and dipping in 2012). Across central Italy, 2002 represents the wettest anomaly ($80\%$ in Rome), whereas 2003 marks severe heatwave drought ($67\%$ in Rome).
		\item \textbf{Dew Point ($T_{\text{dew}}$):} Averages \textbf{$10.73^\circ\text{C}$} in Torrespaccata (peaking at $17.90^\circ\text{C}$ in August, minimum $-1.0^\circ\text{C}$ in Feb-03), \textbf{$10.07^\circ\text{C}$} in Pieve a Nievole, and \textbf{$8.91^\circ\text{C}$} in Montefiascone, with values exceeding the annual baseline during the warm May--October semester across all stations.
		\item \textbf{Barometric Pressure ($P_{\text{surf}}$):} Mean surface pressure is \textbf{$1,015.00\text{ hPa}$} in Pieve a Nievole (max 1992, min 2002), \textbf{$1,014.25\text{ hPa}$} in Montefiascone (min 2012), and \textbf{$1,013.00\text{ hPa}$} in Torrespaccata (May low $1,009\text{ hPa}$, March high $1,017\text{ hPa}$). In Acquapendente, pressure peaked in 2000 and hit a minimum in 2005; in Taviano, pressure peaked in 2004 and dipped in 2010.
	\end{itemize}
	\subsection{Precipitation Regimes: Rain, Snow and Fog}
	All five studies identify a structural shift toward \textbf{short-duration, high-intensity convective rainfall events (``water bombs'')}, increasing localized flash-flood and surface runoff hazards. Simultaneously, \textbf{fog days show a steady multi-decadal decline} linked to rising winter minimum temperatures (Figure~\ref{fig:rain_phenomena} and Figure~\ref{fig:rh_dew_wind}).
	
	\begin{figure*}[t]
		\centering
		\begin{tikzpicture}
			\begin{groupplot}[
				group style={group size=2 by 1, horizontal sep=1.8cm},
				width=0.46\textwidth,
				height=5.6cm,
				ybar,
				symbolic x coords={Acq, Mon, Pie, Tav, Tor},
				xtick=data,
				xticklabels={Acquap., Montef., Pieve N., Taviano, Torresp.},
				xticklabel style={font=\footnotesize, rotate=15, anchor=east},
				ymajorgrids=true,
				grid style={dashed, gray!40},
				nodes near coords,
				every node near coord/.append style={font=\tiny, anchor=south},
				]
				\nextgroupplot[
				title={\textbf{Mean Annual Precipitation (1990--2014)}},
				ylabel={Rainfall (mm/yr)},
				ymin=0, ymax=1800,
				bar width=14pt,
				/pgf/number format/fixed,
				/pgf/number format/precision=1
				]
				\addplot[fill=cyan!75!blue!80] coordinates {(Acq,856.08) (Mon,1503.45) (Pie,1050.10) (Tav,640.15) (Tor,1304.58)};
				
				\nextgroupplot[
				title={\textbf{Cumulative Weather Days (1990--2014)}},
				ylabel={Total Days (25-Yr)},
				ymin=0, ymax=2200,
				bar width=6.5pt,
				legend style={at={(0.5,0.97)}, anchor=north, legend columns=3, font=\scriptsize}
				]
				\addplot[fill=blue!75!black] coordinates {(Acq,1198) (Mon,1281) (Pie,1868) (Tav,1092) (Tor,1711)};
				\addplot[fill=gray!65] coordinates {(Acq,912) (Mon,923) (Pie,1052) (Tav,770) (Tor,254)};
				\addplot[fill=red!75!orange] coordinates {(Acq,181) (Mon,137) (Pie,98) (Tav,188) (Tor,140)};
				\legend{Rain, Fog, Storm}
			\end{groupplot}
		\end{tikzpicture}
		\caption{Mean annual precipitation (left) and 25-year cumulative frequency of rain, fog, and storm days (right) across all five stations.}
		\label{fig:rain_phenomena}
	\end{figure*}
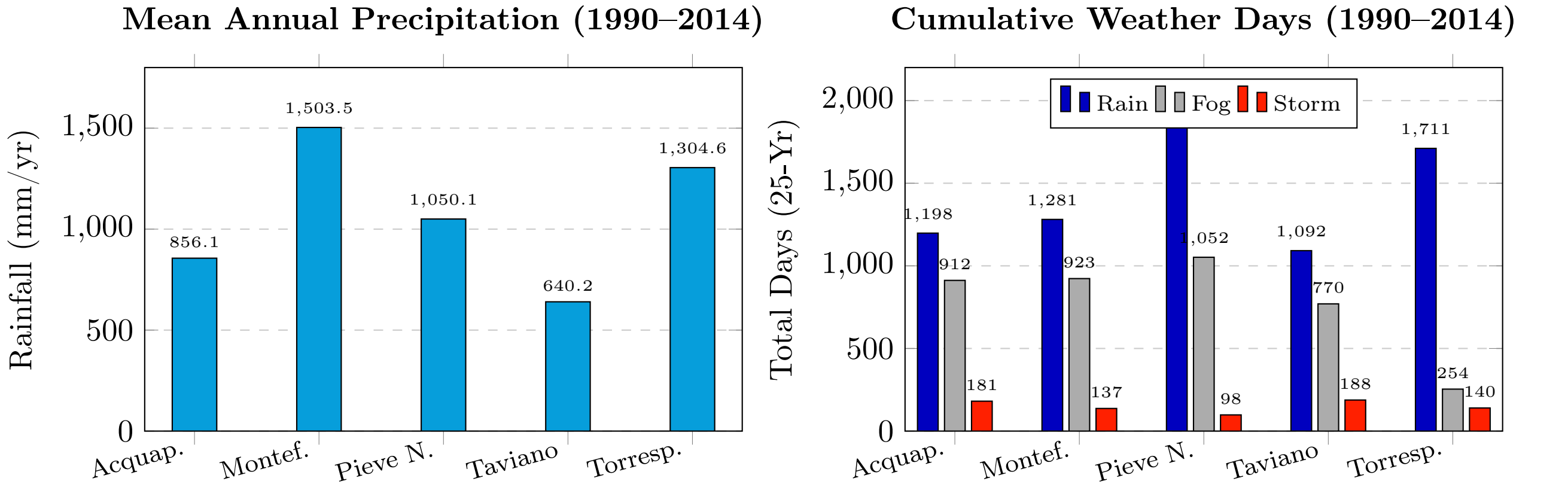
	
	\begin{figure*}[t]
		\centering
		\begin{tikzpicture}
			\begin{groupplot}[
				group style={group size=2 by 1, horizontal sep=1.8cm},
				width=0.46\textwidth,
				height=5.6cm,
				ybar,
				symbolic x coords={Acq, Mon, Pie, Tav, Tor},
				xtick=data,
				xticklabels={Acquap., Montef., Pieve N., Taviano, Torresp.},
				xticklabel style={font=\footnotesize, rotate=15, anchor=east},
				ymajorgrids=true,
				grid style={dashed, gray!40},
				nodes near coords,
				every node near coord/.append style={font=\tiny, anchor=south},
				]
				\nextgroupplot[
				title={\textbf{Mean Annual Relative Humidity (RH \%)}},
				ylabel={Relative Humidity (\%)},
				ymin=0, ymax=90,
				bar width=14pt,
				/pgf/number format/fixed,
				/pgf/number format/precision=2
				]
				\addplot[fill=teal!70!cyan!80] coordinates {(Acq,67.56) (Mon,66.63) (Pie,74.17) (Tav,69.38) (Tor,73.94)};
				
				\nextgroupplot[
				title={\textbf{Mean Dew Point ($^\circ\text{C}$) vs. Daily Wind Speed}},
				ylabel={Value ($^\circ\text{C}$ / km/h)},
				ymin=0, ymax=32,
				bar width=9pt,
				legend style={at={(0.5,0.97)}, anchor=north, legend columns=2, font=\scriptsize},
				/pgf/number format/fixed,
				/pgf/number format/precision=2
				]
				\addplot[fill=blue!65!cyan!80] coordinates {(Acq,8.70) (Mon,8.91) (Pie,10.07) (Tav,11.80) (Tor,10.73)};
				\addplot[fill=orange!75!yellow!85] coordinates {(Acq,23.70) (Mon,18.64) (Pie,21.50) (Tav,24.52) (Tor,22.06)};
				\legend{$T_{\text{dew}}$ ($^\circ\text{C}$), Wind Mean (km/h)}
			\end{groupplot}
		\end{tikzpicture}
		\caption{Comparison of mean annual relative humidity (left), mean dew point temperature, and daily wind speed baseline (right) across the five localities.}
		\label{fig:rh_dew_wind}
	\end{figure*}
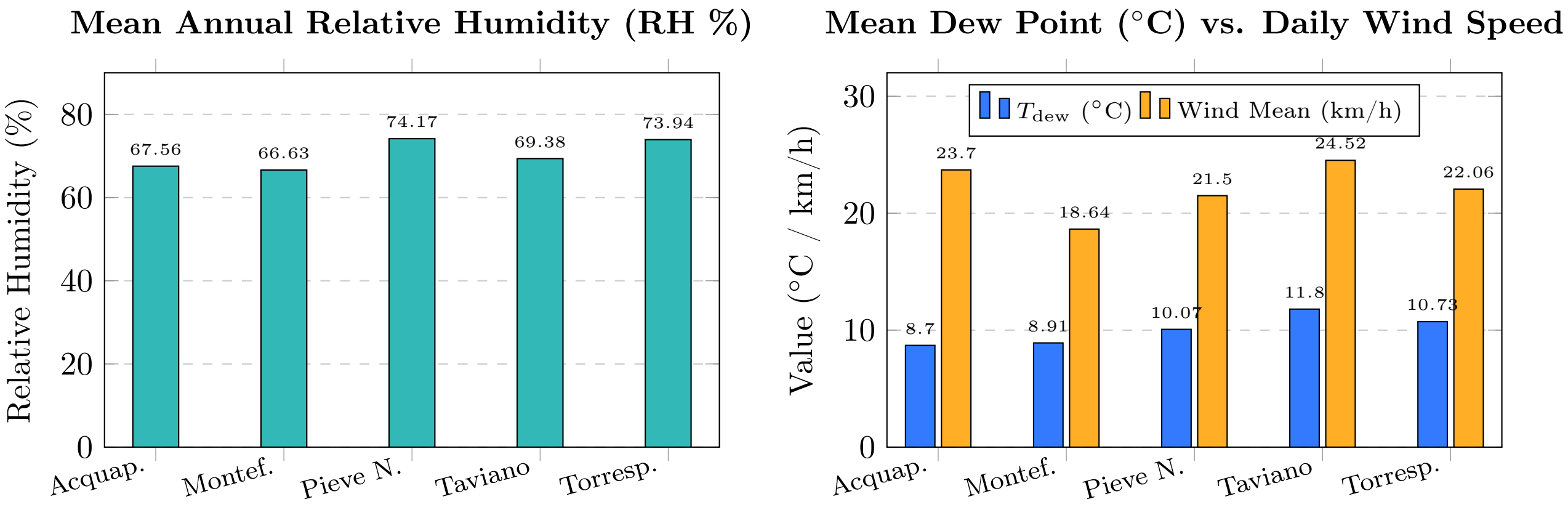
	
	\begin{itemize}
		\item \textbf{Pluviometric Extremes:} Montefiascone records the highest cumulative ($37,586.20\text{ mm}$) and annual mean ($1,503.45\text{ mm/yr}$) rainfall, followed by Torrespaccata ($32,614.60\text{ mm}$, $1,304.58\text{ mm/yr}$), Pieve a Nievole ($26,252.48\text{ mm}$, $1,050.10\text{ mm/yr}$), Acquapendente ($21,401.93\text{ mm}$, $856.08\text{ mm/yr}$), and Taviano ($16,003.70\text{ mm}$, $640.15\text{ mm/yr}$). Wettest annual peaks occur in 2014 (Montefiascone: $3,620\text{ mm}$), 1990 (Pieve a Nievole: $1,844.8\text{ mm}$; Torrespaccata: $2,615\text{ mm}$), 2009 (Acquapendente: $1,413.7\text{ mm}$), and 2004 (Taviano: $1,166.6\text{ mm}$).
		\item \textbf{Lake-Effect and Orographic Snow:} Montefiascone experiences the highest snow frequency ($36\text{ days}$, $588.9\text{ cm}$ cumulative depth, peaking at $115\text{ cm}$ in 2012 and 2013) due to its $590\text{ m}$ elevation and ``lake-effect snow'' from Lake Bolsena. Torrespaccata records $11\text{ snow days}$ (single-event peak $11\text{ cm}$ in Feb-12; seasonal peak $20\text{ cm}$ in 1994), matching Taviano ($11\text{ days}$), while Pieve a Nievole records $9\text{ days}$ and Acquapendente $22\text{ days}$.
		\item \textbf{Fog Distribution:} Radiation fogs peak in October--November in northern/central basins (Acquapendente: $912\text{ days}$; Montefiascone: $923\text{ days}$, $36.92\text{ days/yr}$; Pieve a Nievole: $1,052\text{ days}$) and in January in Taviano ($770\text{ days}$) and Torrespaccata ($254\text{ days}$, $10.2\text{ days/yr}$, dropping from $24\text{ days}$ in 1992 to $8\text{ days}$ in 2014 due to UHI nocturnal warming).
	\end{itemize}
	\vspace{0.2cm}
	\noindent\textbf{Methodological Note on Pluviometric Reconstruction:} 
	The homogenized precipitation series reported in Table~3---specifically the cumulative totals and monthly convective maxima recorded in dense urban sectors such as the Torrespaccata District ($1,304.58\text{ mm/yr}$ mean; $970.00\text{ mm}$ peak in October 1992)---reflect the local microclimatic datasets processed by TS Corporation via statistical imputation to mitigate missing-data gaps~\cite{tasselli2026tor}. Because this reconstruction preserves sensitivity to localized, high-intensity convective cells and urban boundary-layer forcing, cumulative values may exceed standard multi-decadal averages recorded at surrounding open-field synoptic airport stations (e.g., Rome Ciampino).
	
	\subsection{Anemometric Regimes and Local Circulation}
	\begin{itemize}
		\item \textbf{Acquapendente:} Predominant northerly winds; daily mean velocity averages $23.70\text{ km/h}$ (ranging from $9\text{ km/h}$ in September to $21\text{ km/h}$ in July), with autumn/winter gusts reaching \textbf{$36.57\text{ km/h}$} (April 2008).
		\item \textbf{Montefiascone:} Strongly modulated by Lake Bolsena, featuring daytime summer lake breezes (\textit{``Vattuta''}) from the north and intense nocturnal/winter Tramontana and Mistral gales reaching \textbf{$100.00\text{ km/h}$} (March 2001 and September 2014), with a mean daily velocity of $18.64\text{ km/h}$.
		\item \textbf{Pieve a Nievole:} Calm-to-moderate basin regime (mean daily speed $21.50\text{ km/h}$) with monthly minimums of $16\text{--}19\text{ km/h}$ and winter maximums up to \textbf{$33.06\text{ km/h}$} (March 1995).
		\item \textbf{Taviano:} Shielded from western currents by the \textit{Serre Salentine}, while exposed to humid autumn/winter Scirocco (SE) winds driving precipitation; mean daily speed is $24.52\text{ km/h}$, with gusts up to \textbf{$43.45\text{ km/h}$} (March 2002).
		\item \textbf{Torrespaccata:} Governed by a diurnal sea-breeze cycle---nocturnal catabatic drainage from the Colli Albani versus the afternoon Tyrrhenian \textit{``Ponentino''}---with a daily baseline velocity of $22.06\text{ km/h}$ (midrange $19.18\text{ km/h}$) and gusts of \textbf{$50.73\text{ km/h}$} (March 2001) up to \textbf{$86.00\text{ km/h}$} (December 2001).
	\end{itemize}
	
	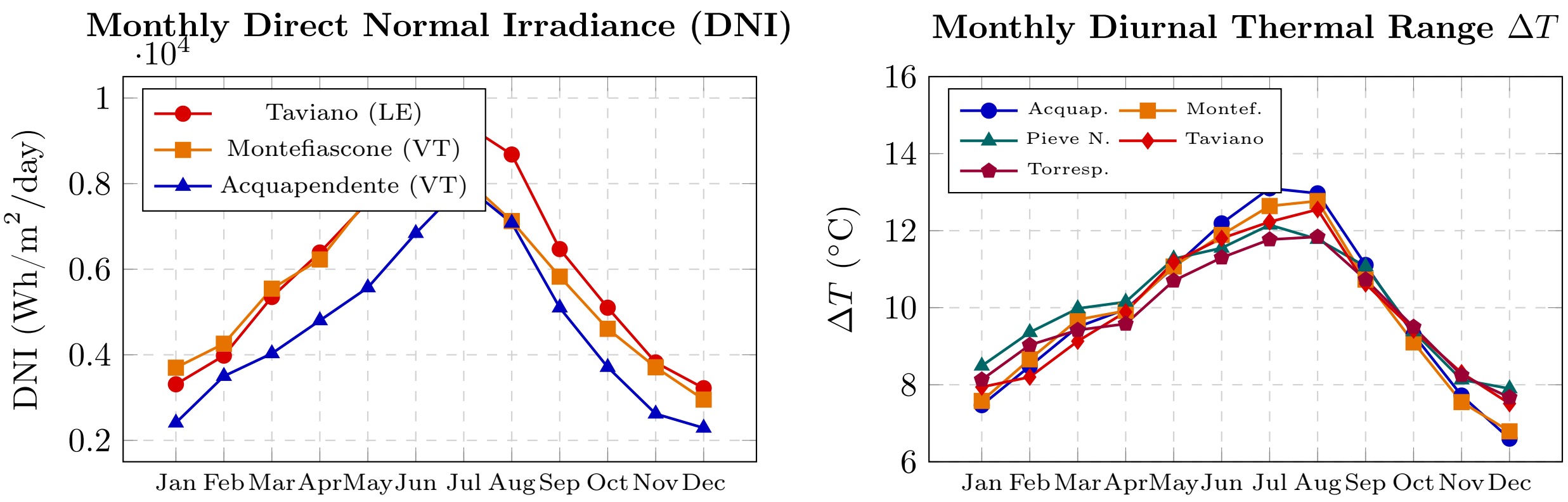
\begin{figure*}[t]
		\centering
		\begin{tikzpicture}
			\begin{groupplot}[
				group style={group size=2 by 1, horizontal sep=1.8cm},
				width=0.46\textwidth,
				height=5.6cm,
				symbolic x coords={Jan,Feb,Mar,Apr,May,Jun,Jul,Aug,Sep,Oct,Nov,Dec},
				xtick=data,
				xticklabel style={font=\scriptsize},
				ymajorgrids=true,
				xmajorgrids=true,
				grid style={dashed, gray!35},
				]
				\nextgroupplot[
				title={\textbf{Monthly Direct Normal Irradiance (DNI)}},
				ylabel={DNI ($\text{Wh/m}^2/\text{day}$)},
				ymin=1500, ymax=10500,
				legend style={at={(0.03,0.97)}, anchor=north west, font=\scriptsize}
				]
				\addplot[color=red!85!black, mark=*, thick] coordinates {(Jan,3310) (Feb,3980) (Mar,5350) (Apr,6390) (May,7570) (Jun,9040) (Jul,9410) (Aug,8680) (Sep,6470) (Oct,5100) (Nov,3820) (Dec,3220)};
				\addplot[color=orange!90!black, mark=square*, thick] coordinates {(Jan,3698.4) (Feb,4259.0) (Mar,5544.6) (Apr,6227.2) (May,7634.2) (Jun,8098.3) (Jul,8152.8) (Aug,7125.5) (Sep,5826.7) (Oct,4606.7) (Nov,3707.4) (Dec,2953.2)};
				\addplot[color=blue!75!black, mark=triangle*, thick] coordinates {(Jan,2410) (Feb,3500) (Mar,4030) (Apr,4800) (May,5570) (Jun,6840) (Jul,7990) (Aug,7080) (Sep,5100) (Oct,3710) (Nov,2620) (Dec,2290)};
				\legend{Taviano (LE), Montefiascone (VT), Acquapendente (VT)}
				
				\nextgroupplot[
				title={\textbf{Monthly Diurnal Thermal Range $\Delta T$}},
				ylabel={$\Delta T$ ($^\circ\text{C}$)},
				ymin=6, ymax=16,
				legend style={at={(0.03,0.97)}, anchor=north west, legend columns=2, font=\tiny}
				]
				\addplot[color=blue!75!black, mark=*, thick] coordinates {(Jan,7.47) (Feb,8.48) (Mar,9.50) (Apr,9.96) (May,11.05) (Jun,12.19) (Jul,13.10) (Aug,12.97) (Sep,11.11) (Oct,9.30) (Nov,7.72) (Dec,6.60)};
				\addplot[color=orange!90!black, mark=square*, thick] coordinates {(Jan,7.58) (Feb,8.66) (Mar,9.69) (Apr,9.92) (May,11.07) (Jun,11.89) (Jul,12.64) (Aug,12.77) (Sep,10.73) (Oct,9.10) (Nov,7.55) (Dec,6.79)};
				\addplot[color=teal!80!black, mark=triangle*, thick] coordinates {(Jan,8.49) (Feb,9.36) (Mar,9.98) (Apr,10.15) (May,11.27) (Jun,11.55) (Jul,12.15) (Aug,11.78) (Sep,11.07) (Oct,9.38) (Nov,8.13) (Dec,7.90)};
				\addplot[color=red!85!black, mark=diamond*, thick] coordinates {(Jan,7.94) (Feb,8.20) (Mar,9.13) (Apr,9.89) (May,11.19) (Jun,11.80) (Jul,12.23) (Aug,12.55) (Sep,10.62) (Oct,9.42) (Nov,8.31) (Dec,7.51)};
				\addplot[color=purple!80!black, mark=pentagon*, thick] coordinates {(Jan,8.13) (Feb,9.03) (Mar,9.42) (Apr,9.58) (May,10.70) (Jun,11.30) (Jul,11.77) (Aug,11.84) (Sep,10.73) (Oct,9.50) (Nov,8.25) (Dec,7.67)};
				\legend{Acquap., Montef., Pieve N., Taviano, Torresp.}
			\end{groupplot}
		\end{tikzpicture}
		\caption{Monthly Direct Normal Irradiance ($\text{Wh/m}^2/\text{day}$) for standard-scale stations (left) and monthly diurnal temperature range $\Delta T$ across all five localities (right).}
		\label{fig:solar_dt}
	\end{figure*}
	
	\begin{table*}[t]
		\centering
		\caption{Monthly mean diurnal temperature range $\Delta T = T_{\text{max}} - T_{\text{min}}$ ($^\circ\text{C}$) across the 1990--2014 baseline.}
		\label{tab:dt_monthly}
		\footnotesize
		\setlength{\tabcolsep}{4pt}
		\resizebox{\textwidth}{!}{%
			\begin{tabular}{lccccccccccccc}
				\toprule
				\textbf{Locality} & \textbf{Jan} & \textbf{Feb} & \textbf{Mar} & \textbf{Apr} & \textbf{May} & \textbf{Jun} & \textbf{Jul} & \textbf{Aug} & \textbf{Sep} & \textbf{Oct} & \textbf{Nov} & \textbf{Dec} & \textbf{Annual Mean} \\
				\midrule
				\textbf{Acquapendente} & $7.47$ & $8.48$ & $9.50$ & $9.96$ & $11.05$ & $12.19$ & $13.10$ & $12.97$ & $11.11$ & $9.30$ & $7.72$ & $6.60$ & $\mathbf{9.97}$ \\
				\textbf{Montefiascone} & $7.58$ & $8.66$ & $9.69$ & $9.92$ & $11.07$ & $11.89$ & $12.64$ & $12.77$ & $10.73$ & $9.10$ & $7.55$ & $6.79$ & $\mathbf{9.87}$ \\
				\textbf{Pieve a Nievole} & $8.49$ & $9.36$ & $9.98$ & $10.15$ & $11.27$ & $11.55$ & $12.15$ & $11.78$ & $11.07$ & $9.38$ & $8.13$ & $7.90$ & $\mathbf{10.10}$ \\
				\textbf{Taviano} & $7.94$ & $8.20$ & $9.13$ & $9.89$ & $11.19$ & $11.80$ & $12.23$ & $12.55$ & $10.62$ & $9.42$ & $8.31$ & $7.51$ & $\mathbf{9.90}$ \\
				\textbf{Torrespaccata} & $8.13$ & $9.03$ & $9.42$ & $9.58$ & $10.70$ & $11.30$ & $11.77$ & $11.84$ & $10.73$ & $9.50$ & $8.25$ & $7.67$ & $\mathbf{9.83}$ \\
				\bottomrule
			\end{tabular}%
		}
	\end{table*}
	\section{Geological, Hydrogeological and Seismological Synthesis}
	
	\subsection{Stratigraphy, Lithology and Geodynamics}
	The five sites represent three distinct Italian geodynamic domains:
	\begin{enumerate}
		\item \textbf{Quaternary Potassic-Ultrapotassic Volcanic Districts (Acquapendente, Montefiascone, Torrespaccata):}
		\begin{itemize}
			\item \textit{Vulsini Volcanic District (Acquapendente \& Montefiascone):} Plio-Pleistocene alkaline-potassic volcanism ($0.8\text{ Ma}$ to $<50\text{ ka}$) covering $\sim 2,200\text{ km}^2$ and resting on $>2,500\text{ m}$ of Pliocene bluish marine clays and Cretaceous-Oligocene \textit{Flysch della Tolfa}. Outcrops include leucitites, basanites, leucititic tephrites, yellow lithoid tuffs, and Montefiascone/Lava Drop ignimbrites. Geochemically, Vulsini rocks show 3-orders-of-magnitude enrichments in radioactive elements ($\text{K, Rb, Th, U}$) as well as Fluorine ($\text{F}$, drinking limit $3\text{ ppm}$) and Arsenic ($\text{As}$), impacting local groundwater quality. Acquapendente also exhibits high geothermal heat flow ($>200\text{ mW/m}^2$), positive gravity/magnetic anomalies, and crustal thinning ($20\text{--}25\text{ km}$).
			\item \textit{Colli Albani / Vulcano Laziale District (Torrespaccata):} Formed by the \textit{Tuscolano-Artemisio} ($0.6\text{--}0.3\text{ Ma}$, $\sim 280\text{ km}^3$), \textit{Faete} ($0.3\text{--}0.2\text{ Ma}$, $\sim 6\text{ km}^3$), and \textit{Hydromagmatic} ($0.2\text{--}0.02\text{ Ma}$, $\sim 1\text{ km}^3$) phases. Pyroclastic covers (\textit{Pozzolane Rosse}, \textit{Tufo Lionato}, \textit{Tufo Rosso a Scorie Nere}, $\sim 0.45\text{ Ma}$) unconformably overlie Middle Pleistocene fluvio-lacustrine deposits (\textit{Valle Giulia Synthem}) and a low-permeability Plio-Pleistocene marine clay aquiclude (\textit{Monte Mario Formation}, $K < 10^{-9}\text{ m/s}$).
		\end{itemize}
		\item \textbf{Northern Apennine Extensional Graben Basin (Pieve a Nievole):} Located in the \textbf{Florence--Prato--Pistoia semi-graben}, controlled by NW--SE normal faults (Prato--Fiesole line). Overthrust Ligurian and Tuscan flyschoid/carbonate units are buried by up to $600\text{ m}$ of Villafranchian fluvio-lacustrine sediments that evolved into the historical \textbf{Fucecchio swamp}, later reclaimed during the Leopoldino period. Slopes exhibit quiescent rototranslative landslides and soil creep.
		\item \textbf{Apulian Carbonate Foreland (Taviano):} Structurally stable karstified carbonate basement (\textit{Calcari di Melissano}) overlain by Plio-Pleistocene \textit{Calcareniti del Salento} (``Tufi'') and \textit{Formazione di Gallipoli} (bluish-gray marly clays and yellowish sandy marls).
	\end{enumerate}
	
	\subsection{Hydrogeology and Anthropic Vulnerability}
	\begin{itemize}
		\item \textbf{Natural Rural Basins:} Acquapendente lies in the Paglia River catchment ($37.5\%$ volcanics, $20\%$ alluvium, $17.5\%$ flysch, $15\%$ clay, $5\%$ debris cones, $2\%$ sands), where fractured lavas and ignimbrites host the primary regional aquifer. Montefiascone exhibits high permeability in lithoid pyroclastics contrasted by low-permeability valleys.
		\item \textbf{Urban Soil Sealing in Torrespaccata:} Situated in the Aniene River sub-basin (Tor Tre Teste / Marrana di Tor Sapienza stream), Torrespaccata suffers from an urban soil sealing index $I_{\text{imp}} > 70\%$. The surface runoff coefficient shifts from $C \approx 0.15\text{--}0.25$ in the unbuilt \textbf{Pratone di Torre Spaccata} (acting as a critical natural retention basin and phreatic aquifer recharge node, $K \approx 10^{-4}\text{--}10^{-5}\text{ m/s}$) to $C \approx 0.85\text{--}0.95$ across built-up areas, amplifying urban flash-flood risk during convective ``water bombs''.
	\end{itemize}
	\subsection{Seismological Comparison (1990--2014)}
\begin{figure*}[t]
		\centering
		\begin{tikzpicture}
			\begin{groupplot}[
				group style={group size=2 by 1, horizontal sep=1.8cm},
				width=0.46\textwidth,
				height=5.6cm,
				ymajorgrids=true,
				grid style={dashed, gray!40},
				]
				\nextgroupplot[
				title={\textbf{Recorded Local Earthquakes per Year (1990--2014)}},
				xlabel={Year},
				ylabel={Number of Events},
				xmin=1989, xmax=2015,
				ymin=0, ymax=95,
				xtick={1990,1994,1998,2002,2006,2010,2014},
				xticklabel style={font=\scriptsize, /pgf/number format/1000 sep={}},
				legend style={at={(0.03,0.97)}, anchor=north west, font=\scriptsize}
				]
				\addplot[ybar, bar width=3.5pt, fill=orange!85!red, draw=none] coordinates {
					(1990,4) (1991,2) (1992,39) (1993,8) (1994,9) (1995,0) (1996,3) (1997,1) (1998,0) (1999,5)
					(2000,7) (2001,4) (2002,0) (2003,0) (2004,4) (2005,3) (2006,32) (2007,18) (2008,75) (2009,43)
					(2010,84) (2011,65) (2012,50) (2013,14) (2014,29)
				};
				\addplot[color=teal!80!black, mark=*, mark size=1.5pt, thick] coordinates {
					(1990,0) (1991,0) (1992,1) (1993,2) (1994,1) (1995,0) (1996,0) (1997,0) (1998,0) (1999,0)
					(2000,3) (2001,0) (2002,1) (2003,0) (2004,0) (2005,1) (2006,2) (2007,0) (2008,0) (2009,1)
					(2010,0) (2011,4) (2012,1) (2013,2) (2014,6)
				};
				\legend{Acquapendente (499), Mon / Pie / Rome (25)}
				
				\nextgroupplot[
				title={\textbf{Historical Peak Wind Speed (1990--2014)}},
				ybar,
				symbolic x coords={Acq, Mon, Pie, Tav, Tor},
				xtick=data,
				xticklabels={Acquap., Montef., Pieve N., Taviano, Torresp.},
				xticklabel style={font=\footnotesize, rotate=15, anchor=east},
				ylabel={Max Wind Gust (km/h)},
				ymin=0, ymax=115,
				bar width=14pt,
				nodes near coords,
				every node near coord/.append style={font=\tiny, anchor=south},
				/pgf/number format/fixed,
				/pgf/number format/precision=2
				]
				\addplot[fill=violet!70!blue!80] coordinates {(Acq,36.57) (Mon,100.00) (Pie,33.06) (Tav,43.45) (Tor,86.00)};
			\end{groupplot}
		\end{tikzpicture}
		\caption{Chronological distribution of local seismic events (left) and maximum recorded wind speed gusts (right) across the 1990--2014 period.}
		\label{fig:seismic_wind}
	\end{figure*}
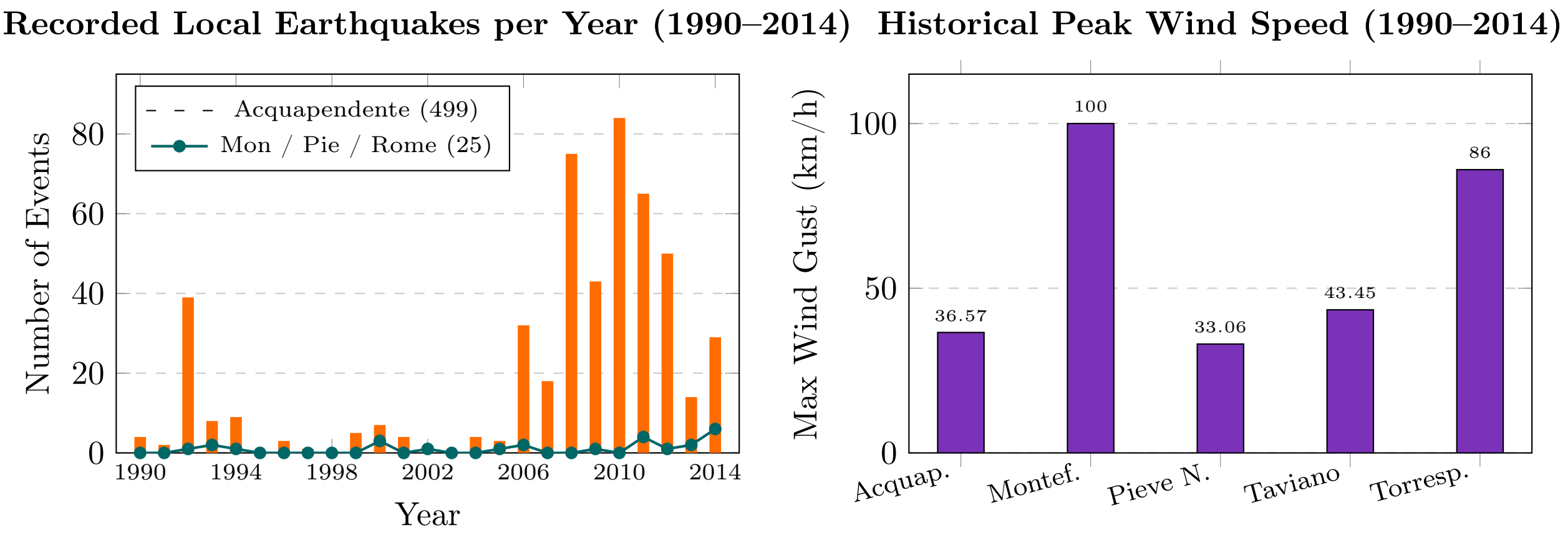

	Seismic activity across all sites is predominantly shallow (hypocentral depths clustered between \textbf{$5\text{ and }10\text{ km}$}), yet frequency and magnitude vary sharply by tectonic setting (Table~\ref{tab:seismic_summary} and Figure~\ref{fig:seismic_wind}).
	
	\begin{table*}[t]
		\centering
		\caption{Recorded local seismicity, magnitude extremes, and hypocentral depth ranges across the five municipalities (1990--2014).}
		\label{tab:seismic_summary}
		\footnotesize
		\setlength{\tabcolsep}{4pt}
		\resizebox{\textwidth}{!}{%
			\begin{tabular}{lccccc}
				\toprule
				\textbf{Seismological Parameter} & \textbf{Acquapendente} & \textbf{Montefiascone} & \textbf{Pieve a Nievole} & \textbf{Taviano} & \textbf{Torrespaccata} \\
				\midrule
				Total Recorded Events (1990--2014) & \textbf{499} & \textbf{25} & \textbf{25} & \textbf{2} & \textbf{25} \\
				Peak Activity Year (Events) & 2010 ($84$ events) & 2014 ($6$ events) & 2014 ($6$ events) & 1999 ($1$), 2009 ($1$) & 2014 ($6$ events) \\
				Minimum Magnitude ($M_w$ / Date) & $0.2$ (2011/12/20) & $0.2$ (2011/12/20) & $1.0$ (2003/06/01) & $0.5$ (2011/12/20) & Low-to-moderate \\
				Maximum Magnitude ($M_w$ / Date) & $3.6$ (2006/07/06) & $3.6$ (1992/07/02) & $2.4$ (2014/10/06) & $2.4$ (2006/07/06) & Moderate regional \\
				Minimum Hypocentral Depth (km) & $0.7$ (2012/08/21) & $0.8$ (2008/06/03) & $3.0$ (2004/10/06) & $5.0$ (2012/08/21) & $\sim 5.0$ \\
				Maximum Hypocentral Depth (km) & $42.5$ (2010/04/28) & $37.3$ (2010/05/18) & $52.0$ (2005/05/13) & $10.0$ (2010/04/28) & $\sim 10.0$ \\
				Dominant Seismogenic Depth (km) & $5.0 - 10.0$ & $5.0 - 10.0$ & $5.0 - 10.0$ & $5.0 - 10.0$ & $5.0 - 10.0$ \\
				\bottomrule
			\end{tabular}%
		}
	\end{table*}
	
	\section{Conclusion}
	Cross-examination of the complete 1990--2014 datasets across Acquapendente, Montefiascone, Pieve a Nievole, Taviano, and Torrespaccata demonstrates a unified regional climatic trajectory modified by local geomorphology and urbanization:
	\begin{enumerate}
		\item \textbf{Mediterraneanization and Thermal Extremes:} The post-2003 acceleration in mean temperatures, summer heatwaves (Aug-03, Jul-12), and intense convective rainstorms indicates a progressive shift toward a sharper wet/dry seasonal contrast across central and southern Italy.
		\item \textbf{Urban vs. Basin Microclimates:} While rural basins (Acquapendente, Pieve a Nievole) remain dominated by nocturnal radiative inversions and persistent autumn fogs ($912\text{--}1,052\text{ days}$), metropolitan fabrics like Torrespaccata exhibit strong Urban Heat Island signatures ($-11.52\text{ HDD/year}$), declining fog days ($254\text{ days}$), and acute flash-flood vulnerability driven by $>70\%$ soil sealing.
		\item \textbf{Territorial Resilience:} Protecting unsealed retention areas (such as \textit{Pratone di Torre Spaccata}), monitoring volcanic groundwater geochemistry ($\text{F, As, Rn/U}$) in the Vulsini district, reinforcing civil infrastructure in active shallow-seismicity sectors (Acquapendente), and shifting from stationary engineering baselines to dynamic climate-adaptive models are imperative for future hazard mitigation.
	\end{enumerate}
	


\begin{thebibliography}{9}
		
		\bibitem{tasselli2015acq}
		Tasselli, D., Ricci, S., \& Bianchi, P. (2015). \textit{New Meteorological and Geological Study of Acquapendente (VT)}. arXiv:1506.04397v1 [astro-ph.IM].
		
		\bibitem{tasselli2015tav}
		Tasselli, D., Ricci, S., \& Bianchi, P. (2015). \textit{New Meteorological and Geological Study of Taviano (LE)}. arXiv:1507.03464v1 [astro-ph.IM].
		
		\bibitem{tasselli2019pie}
		Tasselli, D., Ricci, S., \& Bianchi, P. (2019). \textit{New Astronomical, Meteorological and Geological Study of Pieve a Nievole (PT)}. arXiv:1910.06101v1 [physics.ao-ph].
		
		\bibitem{tasselli2021mon}
		Tasselli, D., Ricci, S., \& Bianchi, P. (2021). \textit{New Astronomical, Meteorological and Geological Study of Montefiascone (VT)}. arXiv:2102.12334v1 [astro-ph.IM].
		
		\bibitem{tasselli2026tor}
		Tasselli, D., Ricci, S., \& Bianchi, P. (2026). \textit{New Astronomical, Meteorological and Geological Study of the ``Torre Spaccata District'' - VII district of the Municipality of Rome}. arXiv:2611.10201v1 [physics.ao-ph].
		
	\end{thebibliography}
\end{document}